# Hydroclimatic time series features at multiple time scales

Georgia Papacharalampous[1,*], Hristos Tyralis[2,3], Yannis Markonis[1], and Martin Hanel[1]

[1] Department of Water Resources and Environmental Modeling, Faculty of Environmental Sciences, Czech University of Life Sciences, Kamýcá 129, Praha-Suchdol 16500, Prague, Czech Republic

[2] Department of Water Resources and Environmental Engineering, School of Civil Engineering, National Technical University of Athens, Heroon Polytechneiou 5, 15780 Zographou, Greece

[3] Construction Agency, Hellenic Air Force, Mesogion Avenue 227–231, 15561 Cholargos, Greece

[*] Correspondence: papacharalampous.georgia@gmail.com, tel: +30 69474 98589

**Email addresses and ORCID profiles:** papacharalampous.georgia@gmail.com, https://orcid.org/0000-0001-5446-954X (Georgia Papacharalampous); montchrister@gmail.com, hristos@itia.ntua.gr, https://orcid.org/0000-0002-8932-4997 (Hristos Tyralis); markonis@fzp.czu.cz, https://orcid.org/0000-0003-0144-8969 (Yannis Markonis); hanel@fzp.czu.cz, https://orcid.org/0000-0001-8317-6711 (Martin Hanel)

**Abstract:** A comprehensive understanding of the behaviours of the various geophysical processes and an effective evaluation of time series (else referred to as "stochastic") simulation models require, among others, detailed investigations across temporal scales. In this work, we propose a novel and detailed methodological framework for advancing and enriching such investigations in a hydroclimatic context. This specific framework is primarily based on a new feature compilation for multi-scale hydroclimatic analyses, and can facilitate largely interpretable feature investigations and comparisons in terms of temporal dependence, temporal variation, "forecastability", lumpiness, stability, nonlinearity (and linearity), trends, spikiness, curvature and seasonality. Multifaceted characterizations are herein obtained by computing the values of the proposed feature compilation across nine temporal resolutions (i.e., the 1-day, 2-day, 3-day, 7-day, 0.5-month, 1-month, 2-month, 3-month and 6-month ones) and three hydroclimatic time series types (i.e., temperature, precipitation and streamflow) for 34-year-long time series records originating from 511 geographical locations across the contiguous United States. Based on the acquired information and knowledge, similarities and differences between the examined time series types with respect to the evolution patterns characterizing their feature values with increasing (or decreasing) temporal resolution are identified. Moreover, the computed features are used as inputs to unsupervised random forests for detecting any meaningful clusters between the examined hydroclimatic time series. This clustering plays an illustrative role within this research, as it facilitates the identification of spatial patterns (with them consisting an important scientific target in hydroclimatic research) and their cross-scale comparison. We find that these specific patterns are largely analogous across temporal resolutions for the examined continental-scale region. We also apply explainable machine learning to compare the features with respect to their usefulness in clustering the time series at the various temporal resolutions. These latter investigations play a vital role within the proposed methodological framework, as they allow interpretation of hydroclimatic similarity at the various temporal resolutions. For most of the features, this usefulness can vary to a notable degree across temporal resolutions and time series types, thereby implying the need for conducting multifaceted time series characterizations for the study of hydroclimatic similarity.

## 1. Introduction

The hydroclimatic time series characteristics are of traditional, fundamental and practical interest to geoscientists (Hurst 1951; Mandelbrot and Wallis 1968; Yevjevich 1974; Moss and Tasker 1987; Montanari 2003; Koutsoyiannis 2013; O'Connell et al. 2016). Numerous scientific questions and targets rotate around these characteristics (see, e.g., the key ones summarized by the initiatives by Montanari et al. 2013; Blöschl et al. 2019), and numerous time series (else referred to as "stochastic") and other statistical models are employed for their exploration and representation (see, e.g., the methods studied and/or reviewed by Carlson et al. 1970; Scheidegger 1970; Moss and Bryson 1974; Yevjevich 1987; Hipel and McLeod 1994; Loaiciga and Leipnik 1996; Montanari et al. 1997; Hamed 2008; Sivakumar and Berndtsson 2010; Lee and Ouarda 2012; Ledvinka 2015; Sivakumar 2017; Slater et al. 2020). The essential role of stochastic models in dealing with hydroclimatic variability and uncertainty has been discussed by Moss and Tasker (1987).

Especially during the latest two decades, hydroclimatic time series characterizations have been notably empowered by the increasing computer capabilities and the release of large hydroclimatic datasets, with many diverse topics being advanced faster than before (Ledvinka and Lamacova 2015; Archfield et al. 2016; Villarini 2016; Blöschl et al. 2017; Do et al. 2017; Hall and Blöschl 2018; Hanel et al. 2018; Berghuijs et al. 2019; Steirou et al. 2019; Papacharalampous and Tyralis 2020). An overview of the large-sample feature investigations in the hydroclimatic research and the research in stochastic (statistical) hydrology can be found in Papacharalampous et al. (2021). In summary, most of these investigations focus on: (a) a single time series characteristic (e.g., autocorrelation, partial autocorrelation, seasonality, trends, temporal variation, entropy, nonlinearity, temporal long-range dependence or extremes); (b) a single time series type (e.g., temperature, precipitation or streamflow); and (c) a single temporal scale (e.g., daily, monthly, seasonal, annual or climatic).

An analysis deviating from point (c) above is usually referred to as "multi-scale" or "cross-scale" analysis. The vital role of such analyses (e.g., the ones by Markonis and Koutsoyiannis 2013; Markonis et al. 2018; McKitrick and Christy 2019; Dimitriadis et al. 2021) towards a comprehensive understanding of hydroclimatic behaviours is widely recognized (see the relevant discussions in McKitrick and Christy 2019), and at the same time various software applications are being developed that study the cross-scale properties of hydroclimatic time series (e.g., those by Hanel et al. 2017; Markonis et al. 2021). Such software tools, as well as their underlying methodologies, are relevant both for increasing our understanding on real-world processes (especially by addressing the scientific question *"How do the characteristics of their time series change with the temporal resolution of the latter increasing or decreasing?"*) through time series analysis and for assessing stochastic simulation methods (see, e.g., the multi-scale time series assessment in Koutsoyiannis and Onof 2001), especially the disaggregation ones, which are fitted to data from two temporal resolutions but still temporal resolutions being in between these two ones might be also of interest from a practical point of view. While some few large-sample hydroclimatic feature investigations diverge from one or two of the above points (a–c) (e.g., those by Markonis et al. 2018; Dimitriadis et al. 2021; Papacharalampous et al. 2021), multi-scale hydroclimatic time series analyses focusing on multiple characteristics and time series types, and related hydroclimatic insights, are currently missing from the literature. Herein, we work towards filling this gap by proposing and thoroughly applying a novel and detailed methodological framework for advancing and enriching multi-scale hydroclimatic investigations.

The various time series models and features appearing across scientific fields are numerous (see, e.g., the thousands of features reviewed and computed by Fulcher et al. 2013; Fulcher and Jones 2014, 2017) and sometimes redundant (Fulcher 2018). Yet, effectively reduced feature compilations for time series analysis exist. Starting from such compilations sourced from the data science literature (Wang et al. 2006; Fulcher et al. 2013; Hyndman et al. 2015; Kang et al. 2017, 2020; Hyndman et al. 2020), and after considering the technical requirements of the problem of interest (i.e., that the feature values should not depend on the time series length and the magnitude-scale of the time series and, additionally, their computational requirements should be low even for the largest time series lengths that appear in hydrology), and after balancing the requirement for interpretability and feature importance information from large-sample investigations on a single temporal resolution (Papacharalampous et al. 2021, 2022), we compose a compilation of features for multi-scale time series analysis in hydroclimatic contexts.

This compilation consists the core of our proposed methodological framework. Its detailed presentation is provided in Section 2.1, while the relevance of its features for conducting hydroclimatic research stems directly from central research themes and concepts appearing in the stochastic (statistical) hydrology field (see again the examples of time series characteristics reported in the point (a) above, as well as the overview, discussions and methodology in Papacharalampous et al. 2021). Focusing on the "multi-scale" dimension of the problem and, at the same time, building on previous efforts for massive time series analyses in the hydroclimatic literature, we use the new feature compilation for multi-scale hydroclimatic investigations as a natural advancement for gaining scientific insights on the topic. This is made by exploiting a large hydroclimatic dataset that comprises temperature, precipitation and streamflow time series. Brief information about these time series is provided in Section 2.2. This latter section further presents the methodological steps followed for comparing the features across temporal resolutions and time series types, together with Section 2.3, while information on the utilized statistical software is provided in Appendix A.

To additionally facilitate the study of spatial hydroclimatic patterns across temporal scales and time series types, the examined geographical locations are algorithmically clustered based on their temperature, precipitation and streamflow features at the various temporal resolutions, as detailed in Section 2.4. These clustering analyses play an illustrative role within the present work, as they facilitate investigations of the spatial variability of the hydroclimatic features, which in general hold a prominent position in hydrology (see, e.g., Kratzert et al. 2019; Jehn et al. 2020). Moreover, explainable machine learning (see, e.g., the reviews on the relevant methods and strategies by Linardatos et al. 2020; Roscher et al. 2020; Belle and Papantonis 2021) through the computation of feature importance scores is exploited for interpreting the clustering outcomes (and hydroclimatic similarity) and the features are ranked according to their usefulness in obtaining the clusters, thereby allowing additional comparisons across temporal resolutions and time series types. These explainable machine learning investigations are of key importance within the proposed methodological framework. The results of the time series analyses and time series clustering are presented in Sections 3.1 and 3.2, respectively. Discussions are also provided in Section 3, while an extended version of the same session can be found in the supplementary material (Appendix B). Furthermore, the most important findings are further discussed in light of the literature in Section 4, where recommendations for future research are also provided. The paper concludes with Section 5.

## 2. Methods and data

### 2.1 Features for multi-scale time series analysis

To approach the problem of interest, we select and compute 23 time series features. These features are briefly documented in the present section (see also the definitions provided in the supplementary material). They include several ones that are based on the sample autocorrelation function (see, e.g., Wei 2006, pp. 18–23), with the most interpretable among them being the lag-1 sample autocorrelation of the time series (`x_acf1`). Another autocorrelation feature selected here, which also summarizes more information than `x_acf1`, is the sum of the squared sample autocorrelation values of the time series at the first ten lags (`x_acf10`). Autocorrelation features that further rely on time series differencing are also selected. Such features are the lag-1 sample autocorrelation of the first-order differenced time series (`diff1_acf1`) and the sum of the squared sample autocorrelation values of the first-order differenced time series at the first ten lags (`diff1_acf10`). Time series differencing is performed by computing the differences between consecutive observations of the time series, and can help to reduce trend and seasonality (Hyndman and Athanasopoulos 2021, Chapter 9.1). First-order differenced data can also be differenced to compute second-order differenced data and so on. Therefore, we also compute the lag-1 sample autocorrelation of the second-order differenced time series (`diff2_acf1`) and the sum of the squared sample autocorrelation values of the second-order differenced time series at the first ten lags (`diff2_acf10`). Furthermore, the magnitude of the relationship between time series values corresponding to the same season of the year in each time series is herein assessed by computing the sample autocorrelation at the first seasonal time lag (`seas_acf1`). For instance, this magnitude is assessed by computing the lag-12 sample autocorrelation and the lag-4 sample autocorrelation for monthly and quarterly data, respectively.

Moreover, the partial autocorrelation function can be used to characterize the correlation between two random variables of a certain phenomenon progressing in time, with the linear dependency between the intervening variables removed (see e.g., Wei 2006, p. 11). Here, we compute the sum of the squared sample partial autocorrelation values of the original time series at the first five lags (`x_pacf5`), the sum of the squared sample partial autocorrelation values for the first five lags of the first-order differenced time series (`diff1x_pacf5`), the sum of the squared sample partial autocorrelation values for the first five lags of the second-order differenced time series (`diff2x_pacf5`) and the sample partial autocorrelation at the first seasonal time lag (`seas_pacf`). The features `x_acf1`, `diff1_acf1`, `diff2_acf1`, `seas_acf1` and `seas_pacf` take values larger than −1 and smaller than 1, and the larger their absolute values the larger the magnitude of the correlation (positive or negative, depending on the sign) between time series data points. The features `x_acf10`, `diff1_acf10`, `diff2_acf10`, `x_pacf5`, `diff1x_pacf5` and `diff2x_pacf5`, on the other hand, take non-negative values, and the larger these values the larger the (partial) autocorrelation magnitude.

Another selected feature whose computation involves time series differencing, similarly to aforementioned features, is the standard deviation of the first-order differenced time series (`std1st_der`). This feature measures the degree of temporal variation in the data, with its larger values indicating larger variation. Furthermore, the spectral entropy of the time series (`entropy`) is computed. This latter feature takes values larger than 0 and smaller than 1, and measures "forecastability" (i.e., the degree in which the future of a time series can be predicted based exclusively on its past; Papacharalampous et al. 2022), with its smaller values indicating larger "forecastability" (Goerg 2013). Its computation involves a univariate normalized spectral density, which is estimated by using an autoregressive (AR) model (for details, see Jung and Gibson 2006).

Two features based on tiled non-overlapping windows are also computed in this work. These features are the lumpiness of the time series (`lumpiness`) and the stability of the time series (`stability`). The length of the tiled windows is set equal to the number of seasons per year (e.g., 12 for monthly time series and 4 for quarterly time series), and their means and variances are computed. Then, the stability is the variance of the means and the lumpiness is the variance of the variances, taking positive and non-negative values, respectively (Kang et al. 2020). Additionally, the nonlinearity of the time series (`nonlinearity`), a modified version of a statistic from the Teräsvirta's nonlinearity test, is computed. This feature takes non-negative values (Kang et al. 2020), with its large values indicating nonlinearity and its values around 0 indicating linearity.

Seven other time series features are computed after applying seasonal and trend decomposition using Loess (STL decomposition; see, e.g., Hyndman and Athanasopoulos 2021, Chapter 3.6) to the time series. This application is herein made according to the procedures described in Hyndman and Khandakar (2008), and the seasonal, smoothed trend and remainder components of the time series are obtained. Then, the trend strength of the time series (`trend`) is defined as 1 minus the quotient of the variance of the remainder component (sometimes referred to as the "random" component; Cleveland et al. 1990) and the variance of the time series with its seasonal component removed (Wang et al. 2006). This feature takes values larger than or equal to 0 and smaller than 1, with its smaller values indicating weaker trends.

The strengths of spikes (which, according to Goin and Ahern 2019, are acute increases in the time series followed by immediate returns to the underlying level of the time series; `spike`), linearity (`linearity`) and curvature (`curvature`) of the time series are also computed after applying STL decomposition and by taking into consideration measures of trend and seasonality of the time series. More precisely, the strength of spikes (or spikiness) of a time series is computed as the variance of the leave-one-out variances of its remainder component and takes values from 0 to 1 (Kang et al. 2020). On the other hand, the linearity and curvature measures are computed based on the primarily coefficients of an orthogonal quadratic regression and take values from −∞ to +∞ (Kang et al. 2020). Note also that the feature `linearity` is not identical (or approximately identical) and should not be confused with the feature `nonlinearity` (see above). Two other features computed for the remainder component of the time series are its lag-1 sample autocorrelation (`e_acf1`) and the sum of its squared sample autocorrelation values at the first ten time lags (`e_acf10`). The former feature takes values from −1 to 1, and the latter one takes values larger than 0. Similarly to the trend strength of the time series, its seasonality strength (`seasonal_strength`) is defined as 1 minus the quotient of the variance of the remainder component and the variance of the time series with its smoothed trend component removed (Wang et al. 2006). It takes values from 0 to 1, with its smaller values indicating weaker seasonality.

## 2.2 Hydroclimatic data at multiple time scales

To explore the values of the selected time series features (see Section 2.1) across multiple time scales and hydroclimatic time series types, we exploit information included in the CAMELS (Catchment Attributes and Meteorology for Large-sample Studies) dataset. This dataset has been made available by Newman et al. (2014) and Addor et al. (2017a), and has already facilitated a variety of large-sample investigations (see, e.g., those by Dimitriadis et al. 2021; Tyralis et al. 2021a). Documentations of this dataset can be found in Newman et al. (2015) and Addor et al. (2017b). In brief, the entire dataset comprises, among others, daily hydrometeorological and streamflow time series from 671 small- to medium-sized catchments across the contiguous United States. The hydrometeorological time series have been obtained by processing the dataset by Thornton et al. (2014), and include minimum temperature, maximum temperature and precipitation time series.

Herein, we specifically focus on the 511 geographical locations (Figure 1), for which complete daily time series are available for the 34-year period between 1980 and 2013. Indeed, time series gaps would have to be filled before computing the selected time series features, with the gap filling modelling procedures possibly affecting the feature values to some extent. For the geographical locations of focus, we compute complete 34-year-long time series of daily temperature means by averaging the available daily minimum and maximum time series values. These new time series constitute the daily dataset examined in this work, together with the complete 34-year-long daily precipitation and streamflow time series corresponding to the same geographical locations. Subsequently, we extend our dataset by computing temperature, precipitation and streamflow time series of 2-day, 3-day, 7-day, 0.5-month, 1-month, 2-month, 3-month and 6-month temporal resolutions. This is made by aggregating the temperature, precipitation and streamflow time series of 1-day temporal resolution. The extended dataset includes 511 (number of the examined geographical locations) × 3 (number of the examined time series types) × 9 (number of the examined temporal resolutions) = 13 797 time series of varying lengths.

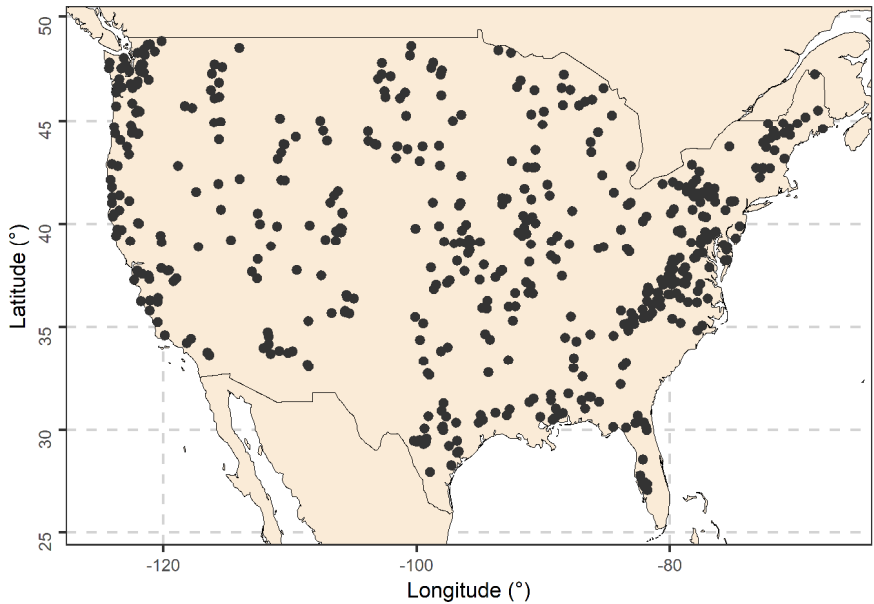

Figure 1. Examined geographical locations across the contiguous United States.

## 2.3   Feature computation and summaries

We (i) scale each time series of the extended dataset (see Section 2.2) to mean 0 and standard deviation 1, and (ii) summarize it by computing its features (see Section 2.1). With step (i), we assure that the computed feature values do not depend on the scale (i.e., the magnitude) of the time series. Step (ii) is made by assuming seasonal behaviour and predefined number of seasons per year. The latter parameter is set to 365, 182, 121, 52, 24, 12, 6, 4 and 2 for the 1-day, 2-day, 3-day, 7-day, 0.5-month, 1-month, 2-month, 3-month and 6-month temporal resolutions, respectively. The computed feature values are 13 797 (number of time series) × 23 (number of features) = 317 331 and are summarized with boxplots and mean values. These summaries are made separately for each set {time series type, temporal resolution, feature} and facilitate the targeted comparisons of the feature values across temporal resolutions, as well as the targeted comparisons of the time series types with respect to the evolution patterns characterizing their features with increasing (or decreasing) temporal resolution.

## 2.4 Clustering and spatial explorations

Separately for each set {time series type, temporal resolution}, we use random forests (Breiman 2001) in unsupervised mode (and without performing cross validation) for clustering the time series (and, by extension, their corresponding geographical locations) into four groups (with this number being selected a priori as a reasonable one considering the size of the region under investigation and retained common in the various cluster analyses to facilitate direct comparisons across time scales and time series types) based on the feature values. The latter are simply the random forests' inputs, while spatial information is not considered in the clustering (which is solely feature-based). The same applies for information on the magnitude of the time series, as the feature values do not depend on this magnitude. Random forests are fitted with 5 000 trees (which are considered as a large number of trees, in general), as: (a) their performance is expected to be adequate when the number of trees exceeds (or is equal to approximately) 500 and known to increase with increasing the number of trees (see, e.g., the review by Tyralis et al. 2019a); and (b) the utilized PC can afford the computational requirements that are imposed by this methodological choice. The number of variables randomly sampled as candidates at each split is set to 2 and the remaining parameters (e.g., the minimum size of terminal nodes) are kept as default, as random forests can offer satisfactory results even without optimization as long as they are run with a large number of trees (Tyralis et al. 2019a). The obtained clusters of geographical locations are depicted in spatial visualizations, which allow the targeted comparisons of spatial hydroclimatic patterns across temporal resolutions and time series types.

Although the clustering analyses play only an illustrative role in this work, it is still relevant to assess the reliability of their results. This is made by computing silhouette widths (Rousseeuw 1987) and their averages. Silhouette widths show "which objects lie well within their cluster and which ones are merely somewhere in between clusters" (Rousseeuw 1987). More precisely, objects with a silhouette width almost equal to 1 are very well clustered, while a silhouette width almost equal to 0 means that the object lies between two clusters and objects with a negative silhouette width are probably placed in the wrong cluster. Obviously, similarly to what applies to each individual silhouette width, the larger the average silhouette width of a specific cluster analysis, the more reliable this analysis.

Aside from computing the average silhouette widths for the clustering analyses that divide the dataset into four clusters, we also compute them for the variants of the same analyses that divide the dataset into two, three, five, six, seven, eight, nine and ten clusters. This is made for finding the optimal numbers of clusters in terms of average silhouette width for the various sets {time series type, temporal resolution} and for the exploited dataset. Then, these optimal numbers of clusters facilitate additional clustering analyses (which again rely on the use of random forests with 5 000 trees) and additional comparisons across time series types and temporal resolutions. Notably, the number of clusters can be considered as a hyperparameter in clustering; thus, it is accurate to state that the additional clustering analyses (i.e., those dividing the datasets into the optimal numbers of clusters) involve hyperparameter optimization (while those dividing the datasets into the a priori selected number of clusters do not).

Random forests are also exploited for obtaining rankings of the features according to their usefulness-importance in each time series clustering setting, and comparisons of these rankings are performed across temporal scales and across time series types. It should be noted, at this point, that ideally we wish to know the contribution of each feature into explaining the phenomena under study in the form of percentages instead of only knowing the corresponding rankings. Still, according to Boulesteix et al. (2012), these rankings consist the only meaningful information that we can obtain through the computation of variable importance scores. Thus, after having computed scores of this type, we use them to rank the features and present the rankings. Specifically, we compute the "mean decrease Gini" variable importance metric, which is the total decrease in node impurities from splitting on the variable, as measured by the Gini index and averaged over all the trees (Liaw 2018). Perhaps it is also relevant to note here that the computation of this metric does not rely on (and is independent of) the selection of the number of clusters. The computation of feature importance scores using random forests is documented as a "random forest explainability" approach and an "explainable machine learning" approach, for instance, in the review by Belle and Papantonis (2021, Section 8.1.2).

For detailed information on the properties and the utilities of random forests, the reader is referred to the review by Tyralis et al. (2019a). In summary, the core idea behind their utilization for clustering is the following: Real cases (with each having been summarized with a set of feature values) that are similar to each other will frequently end up in the same terminal node of a tree. The percentage of the time at which they actually end up in the same terminal node of a tree is measured by computing the proximity matrix. Therefore, the proximity matrix can be used as a similarity measure, and clustering using this similarity can be performed to divide the original real cases into groups (Liaw and Wiener 2002).

## 3. Results and discussions

### 3.1 Hydroclimatic time series characterizations

In Figures 2–5, we present the side-by-side boxplots of the computed feature values for the temperature, precipitation and streamflow time series at the nine examined temporal resolutions (with each boxplot summarizing the values of a specific feature for a specific set {time series type, temporal resolution}), while in Figure 6 we present the means of these feature values. To ease the extraction of information required for approaching the problem of interest, in this latter figure we also report the rankings of the mean values from the smallest ($1^{st}$) to the largest ($9^{th}$). Overall, distinct patterns are found to characterize the feature values as we move from the time series with the finest temporal scale to those with the coarser temporal scale. These patterns differ, to a larger or smaller extent, for the various time series features and for the three time series types.

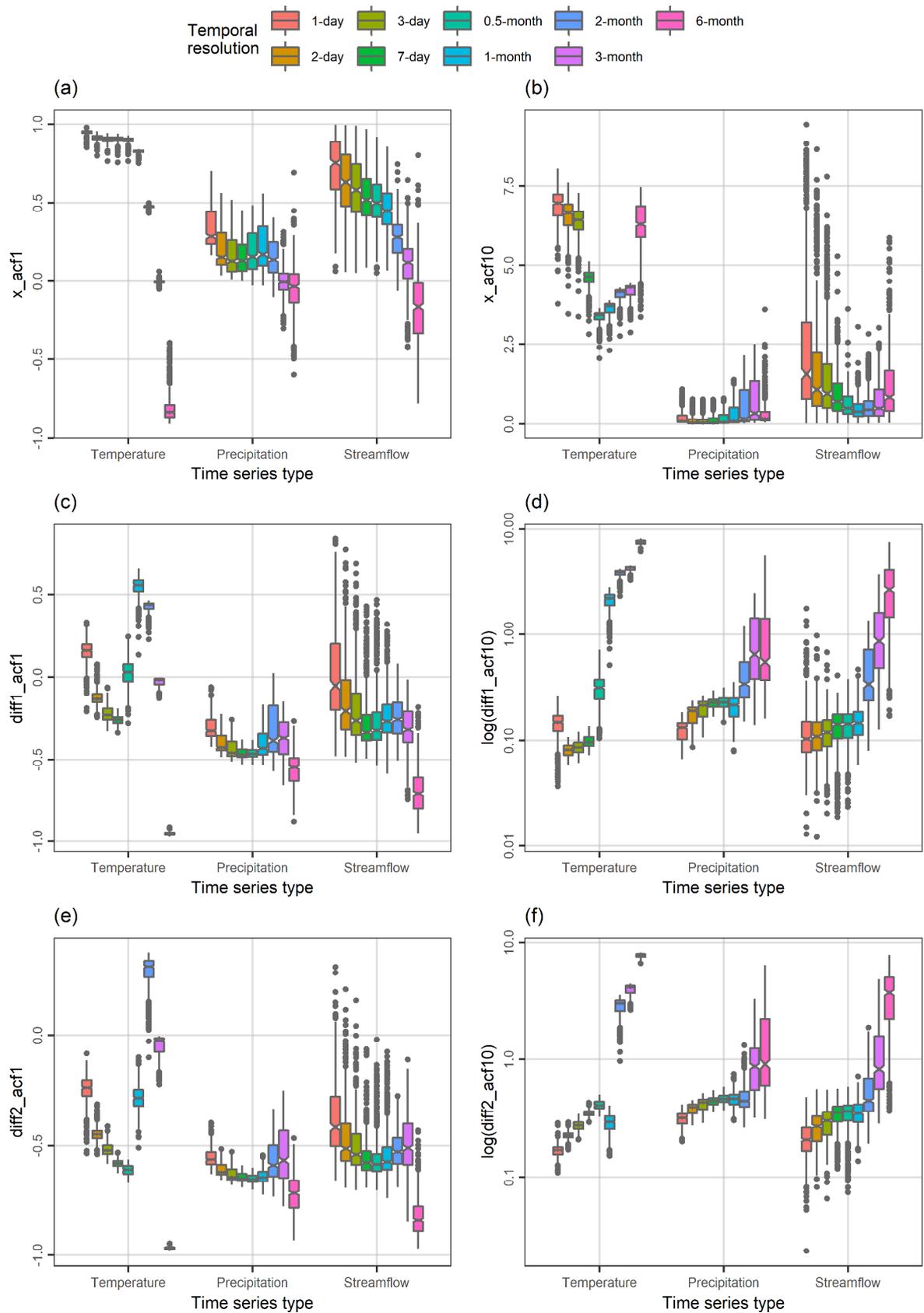

Figure 2. Side-by-side boxplots of the values of the features (a) `x_acf1`, (b) `x_acf10`, (c) `diff1_acf1`, (d) `diff1_acf10`, (e) `diff2_acf1` and (f) `diff2_acf10` computed for the temperature, precipitation and streamflow time series at the various examined temporal resolutions. The features are defined in Section 2.1.

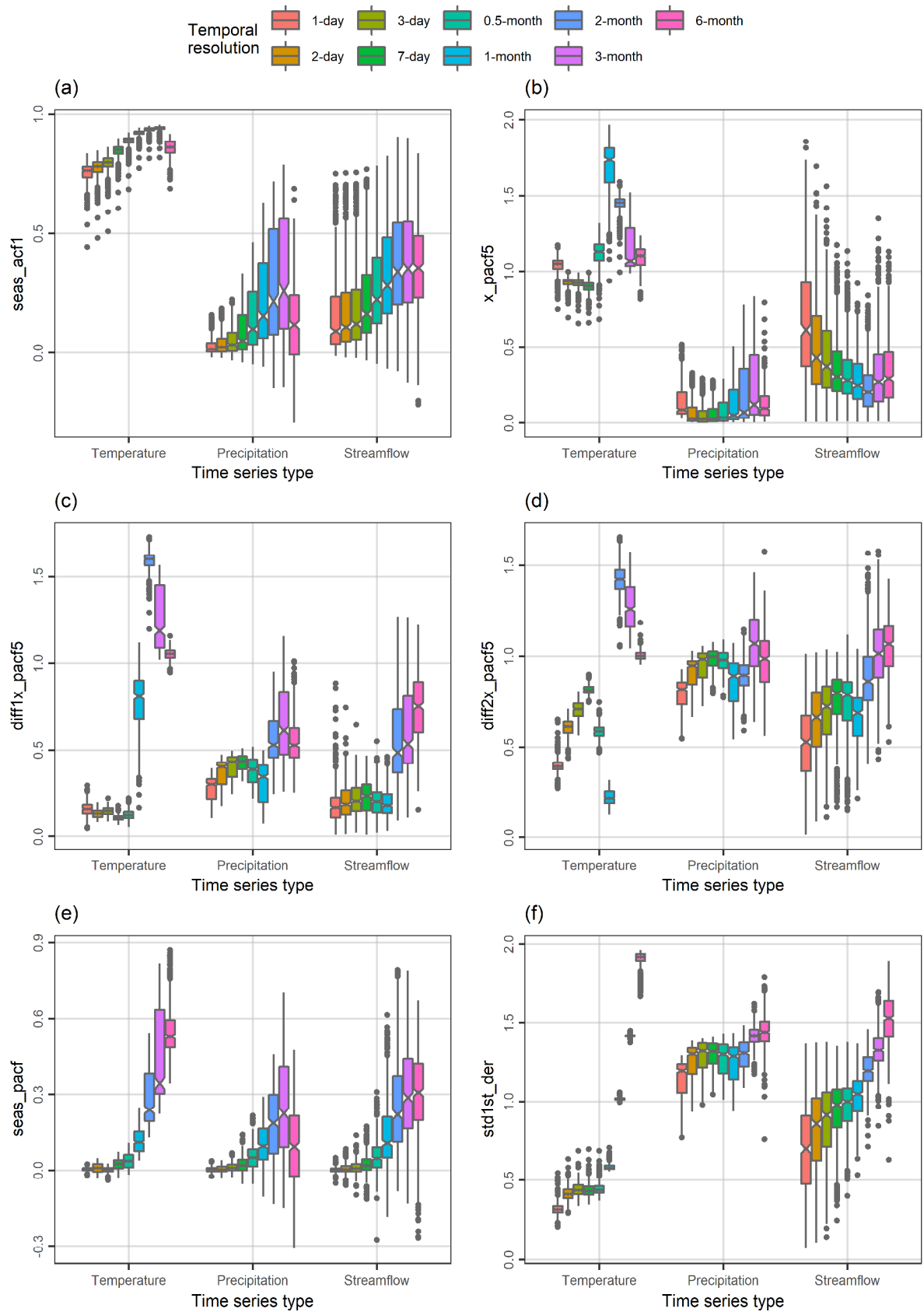

Figure 3. Side-by-side boxplots of the values of the features (a) `seas_acf1`, (b) `x_pacf5`, (c) `diff1_pacf5`, (d) `diff2_pacf5`, (e) `seas_pacf` and (f) `std1st_der` computed for the temperature, precipitation and streamflow time series at the various examined temporal resolutions. The features are defined in Section 2.1.

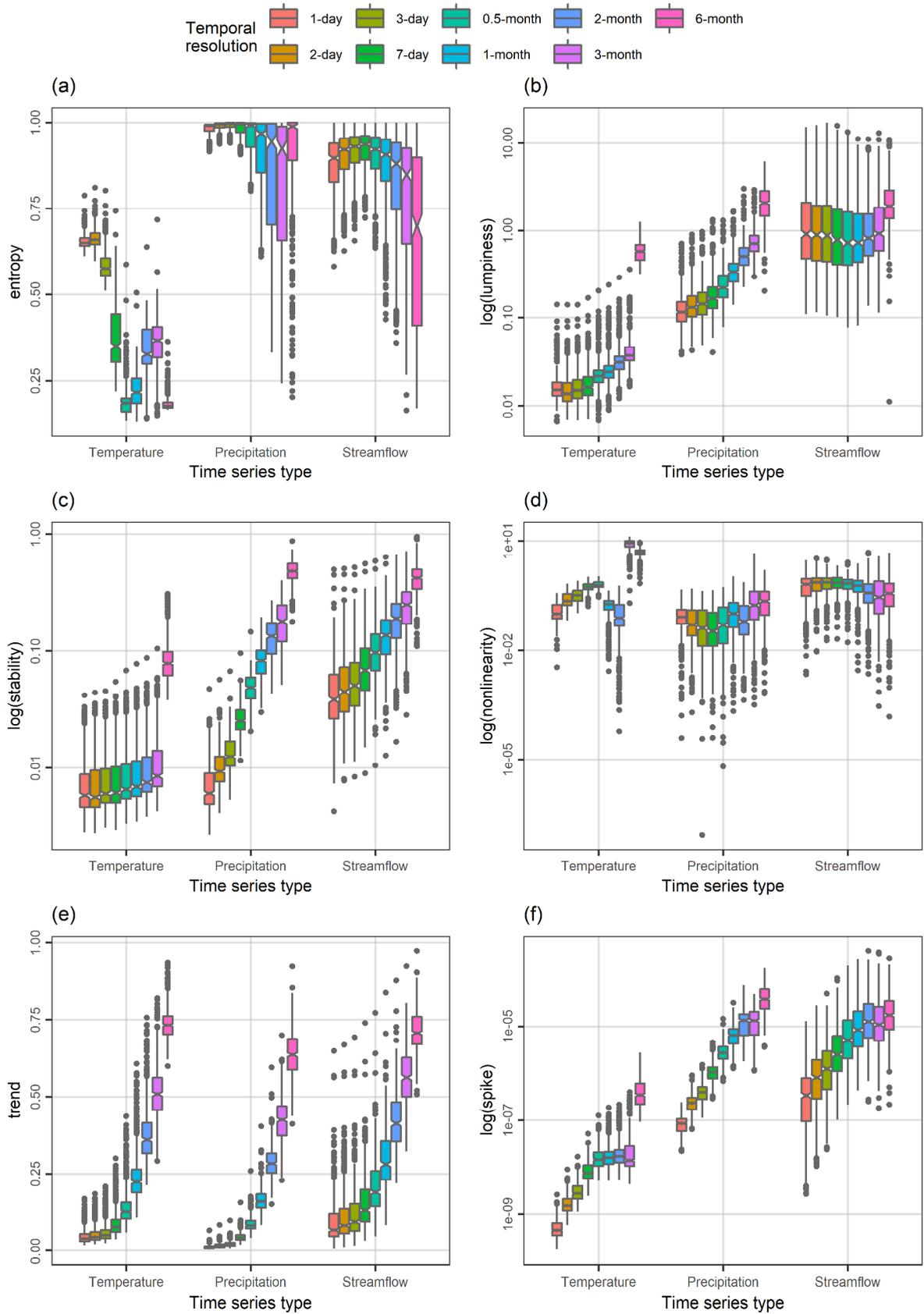

Figure 4. Side-by-side boxplots of the values of the features (a) `entropy`, (b) `lumpiness`, (c) `stability`, (d) `nonlinearity`, (e) `trend` and (f) `spike` computed for the temperature, precipitation and streamflow time series at the various examined temporal resolutions. The features are defined in Section 2.1.

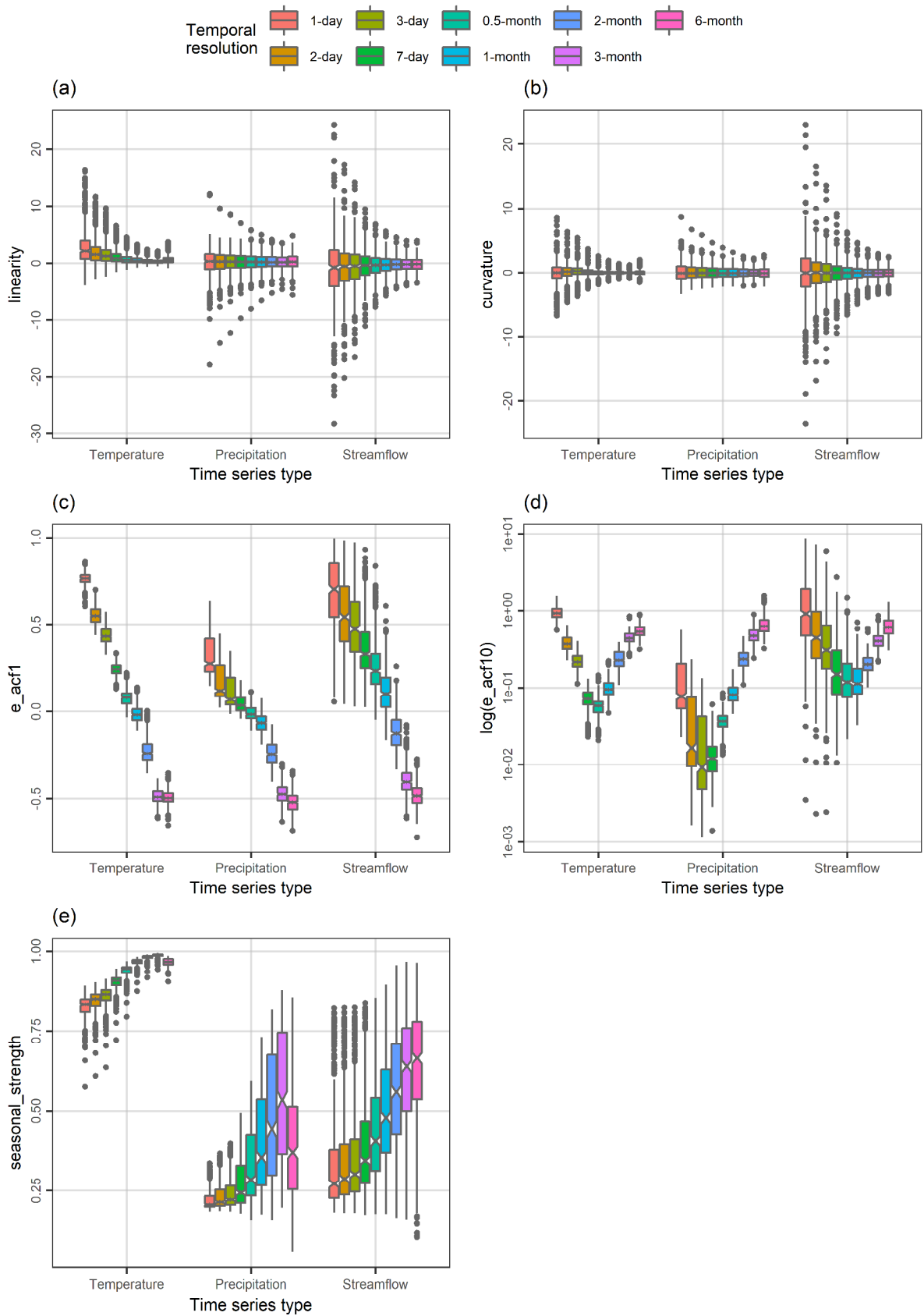

Figure 5. Side-by-side boxplots of the values of the features (a) `linearity`, (b) `curvature`, (c) `e_acf1`, (d) `e_acf10` and (e) `seasonal_strength` computed for the temperature, precipitation and streamflow time series at the various examined temporal resolutions. The features are defined in Section 2.1.

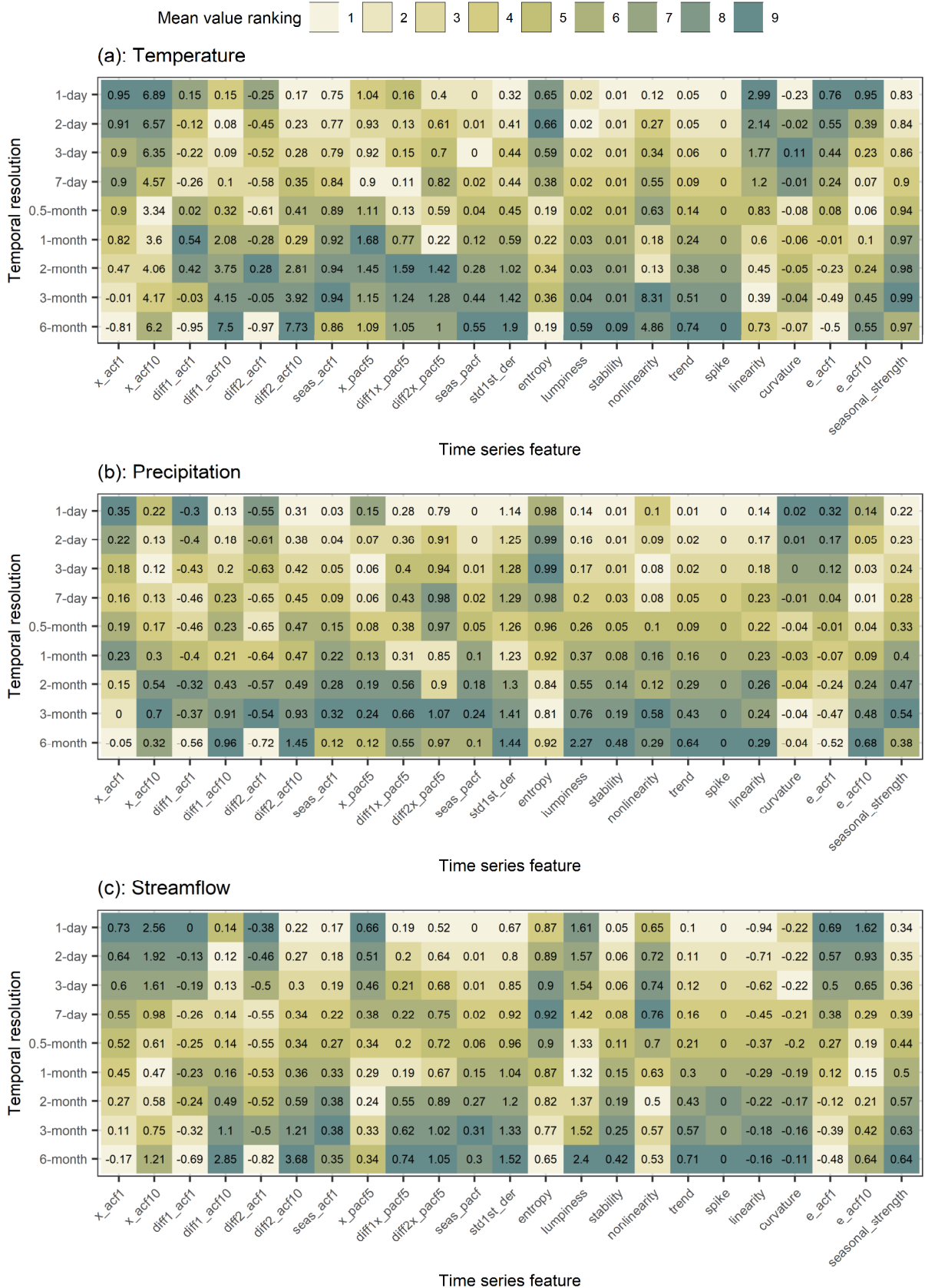

Figure 6. Mean values of the features of the (a) temperature, (b) precipitation and (c) streamflow time series at the various examined temporal resolutions, and their rankings from the smallest (1st) to the largest (9th). The features are defined in Section 2.1. Features that appear to be zero (e.g., `spike`) take very low values.

More precisely, the lag-1 sample autocorrelation (`x_acf1`) mostly becomes weaker as we move from the finest (i.e., the daily) temporal resolution to the coarser ones up to the 3-month temporal resolution (in which the lag-1 sample autocorrelation becomes either close to zero, indicating almost uncorrelated variables, or weak positive, with both these cases being, most probably, largely related to the remarkably strong seasonality features at this specific temporal resolution, due to which all the pairs of consecutive values in each time series could be less correlated than in other temporal resolutions; see Figures 3a and 5e) for all the three examined time series types (see Figure 2a). This pattern of decreasing lag-1 sample autocorrelation for the above-reported time scales seems reasonable, as two days in the row are rather expected to be more similar than two months in the row or two 3-month seasons in the row for the continental-scale region under investigation. Still, this same pattern is somewhat more pronounced for the temperature and streamflow time series.

Also, the precipitation time series are, in general, characterized by notably smaller lag-1 sample autocorrelation values than the temperature and streamflow time series, indicating their largely expected smaller degree of persistence (especially compared to the temperature time series). For the 6-month temporal resolution, the lag-1 sample autocorrelation is more intense than for the 3-month temporal resolution and (for most, if not all, the stations) negative for all the three examined time series types. Especially for the temperature time series, the same feature takes remarkably large absolute values, which could be explained, to some extent, based on our current process understanding. Indeed, temperatures are known to have a strong annual periodicity with low negative values during winter and high positive values during summer for many parts of the contiguous United States.

Furthermore, the sum of the squared sample autocorrelation values at the first ten time lags (`x_acf10`) is characterized by different patterns, with the largest values tending to characterize the finest and coarsest temporal scales, and medium values tending to characterize the remaining temporal scales, especially for the temperature and streamflow time series (see Figure 2b). Indeed, the dependence in these time series at the 1-month temporal resolution is rather expected to be less intense, not only for the first time lag (i.e., for the consecutive values) but for some other time lags (e.g., the second, third, etc.) as well, than the persistence at the 1-day and 6-month temporal resolutions at the same time lags (thereby leading to smaller sums of sample autocorrelation values at the former temporal resolution than at the latter ones). As it is rather expected, the same pattern is less pronounced for the case of precipitation, for which persistence is notably less intense. In fact, a sound exception to the above-outlined rule is observed for the precipitation time series at the 6-month temporal resolution. Indeed, the `x_acf10` values of these time series are, on average, less intense than their respective values at the 2-month and 3-month temporal resolutions (see Figure 6b).

In its turn, time series differencing leads to different autocorrelation patterns across temporal resolutions and time series types (see Figure 2c–f). The most pronounced among these patterns are observed for the sum of the squared sample autocorrelation values of the first-order differenced time series at the first ten lags (`diff1_acf10`), as well as those observed for the sum of the squared sample autocorrelation values of the second-order differenced time series at the first ten lags (`diff2_acf10`). As time series differencing is expected to weaken the trend and seasonality features in the time series, patterns observed for the autocorrelation features `diff1_acf1`, `diff1_acf10`, `diff2_acf1` and `diff2_acf10` could be particularly interesting to discuss. For instance, for the temperature time series, the `diff1_acf1` and `diff2_acf1` values follow a completely different pattern compared to the `x_acf1` values, while the same does not hold for the precipitation and streamflow time series (see Figure 2a,c,e). This could be partly expected due to the more pronounced seasonality features in the temperature time series than in the precipitation and streamflow time series (see also the relevant quantifications of this work in Figures 3a and 5e, and the related discussions), although relevant expectations do not exist a priori for the trend characteristics.

The somewhat expected patterns of increasing seasonality as we move from the finest temporal scales to the coarser ones up to the 3-month temporal resolution are confirmed by the computed values of the sample autocorrelation at the first seasonal time lag (`seas_acf1`) (see Figure 3a). On the other hand, the sample autocorrelation values at the first seasonal time lag are, more or less, smaller for the 6-month temporal resolution than for some of the finer temporal resolutions. Notably, the fact that the largest sample autocorrelation value at the first seasonal time lag appears at the 3-month temporal resolution seems reasonable, although related proofs are not available in the literature. Moreover, the patterns identified across temporal resolutions and time series types for the four partial autocorrelation features (i.e., `x_pacf5`, `diff1_pacf5`, `diff2_pacf5` and `seas_pacf`) are, to some extent, similar to patterns identified for autocorrelation features (see Figures 2, 3b–e and 6). Furthermore, the standard deviation of the first-order differenced time series (`std1st_der`), computed after scaling the time series to mean 0 and standard deviation 1, gets larger on average with increasing temporal scale (see Figures 3f and 6), indicating a larger degree of temporal variation other than that associated to trend and seasonal features when the temporal scale increases.

Spectral entropy (`entropy`) is characterized by unique patterns, which are also quite different for temperature than they are for precipitation and streamflow (see Figure 4a). Based on the means of the computed `entropy` values (see Figure 6), we could rank the examined temporal resolutions from the most to the least "forecastable" ones, as follows for temperature: 6-month, 0.5-month, 1-month, 2-month, 3-month, 7-day, 3-day, 1-day and 2-day temporal resolutions. The respective order for the precipitation time series is: 3-month, 2-month, 6-month, 1-month, 0.5-month, 1-day, 7-day, 2-day and 3-day temporal resolutions. Similar (but not identical) patterns are extracted for the streamflow time series, which could be ordered from the most to the least "forecastable" ones as follows: time series of 6-month, 3-month, 2-month, 1-day, 1-month, 2-day, 0.5-month, 3-day and 7-day temporal resolutions. Notably, the fact that the computed `entropy` values are, at least on average, larger at the finest temporal resolutions (indicating lower forecastability at these temporal resolutions), especially for the temperature and precipitation time series, could be largely expected. Indeed, as the seasonal patterns become more concrete, forecastability is rather expected to increase.

Although such an association unarguably exists, at least to some extent, it seems that the patterns between the `entropy` feature and each of the `seas_acf1` and `seasonal_strength` features are not analogous. On the other hand, large similarities are observed in the patterns across temporal resolutions and time series types for the lumpiness (`lumpiness`) and the stability (`stability`) of the time series, with these feature values mostly increasing with increasing temporal scale (see Figure 4b,c), indicating that the variance and mean of the time series are more variable across its different parts with such an increase. The only exceptions to this pattern are observed for the `lumpiness` values computed for streamflow, whose mean values for the various temporal resolution (apart from the 6-month temporal resolution) are mostly very close to each other (see Figure 6).

The temperature, precipitation and streamflow time series exhibit quite varying patterns in terms of nonlinearity (`nonlinearity`) across temporal resolutions (see Figures 4d and 6). Interesting motives are also identified for the trend strength (`trend`) and the spikiness (`spike`) of the time series (see Figures 4e,f and 6), whose values (mostly) increase with increasing temporal scale, while two features exhibiting very different patterns than the remaining ones are the linearity (`linearity`) and curvature (`curvature`) of the time series (see Figures 5a, b and 6). Furthermore, distinct patterns are also identified for the lag-1 sample autocorrelation (`e_acf1`) and the sum of the squared sample autocorrelation values at the first ten lags (`e_acf10`) of the remainder component of the time series (see Figures 5c,d and 6).

Notably, these patterns somewhat resemble the patterns obtained for the lag-1 sample autocorrelation of the original time series (`x_acf1`) and the sum of the squared sample autocorrelation values at the first ten time lags on the original time series (`x_acf10`), as it could be expected (at least to some extent) based on the corresponding feature definitions, but they are still more concrete than them (due to the removal of the trend-cycle and seasonality components, as these components are identified by the time series decomposition algorithm). Also notably, the patterns extracted from the computed values of the strength of seasonality (`seasonal_strength`) in the time series (see Figures 5e and 6) are similar to those extracted from the `seas_acf1` values (see again Figures 3a and 6), thereby further confirming our a priori expectations on an increasing intensity of the seasonality patterns as the temporal scale becomes coarser up to the 3-month temporal resolution for the temperature and precipitation time series and up to the 6- month temporal resolution for the streamflow time series.

Additional similarities and differences to those outlined above emerge from the results (see, e.g., Figure 6). Indeed, the features `std1st_der` and `entropy` are found to have opposing patterns as regards their changes with increasing (or decreasing) temporal resolution, especially for temperature and precipitation, but for streamflow as well. Other examples of feature pairs with opposing patterns are the following ones: {`e_acf10`, `seasonal_strength`} (mostly for streamflow), {`entropy`, `nonlinearity`} (for temperature), {`x_acf1`, `nonlinearity`} (for temperature) and {`diff1_acf10`, `e_acf1`} (for precipitation). Moreover, pairs with quite similar patterns (again as regards their changes with increasing or decreasing temporal resolution) that have not been discussed in the above paragraphs are the following ones: {`curvature`, `e_acf1`} (for precipitation), {`trend`, `seasonal_trend`} (for streamflow) and {`trend`, `linearity`} (for precipitation and streamflow).

It should, lastly, be noted that the features appearing to be zero on Figure 6 take very low values, in reality. The most representative case is that of the feature `spike` (which appears to be zero at all the time series and time series types); yet, other features (i.e., the `x_acf1`, `diff1_acf1`, `seas_pacf` ones) appear to be zero for specific time series type(s) and temporal resolution(s) as well. To perceive the range of values that these features take at the various time scales and for the various time series types, as well as how the very low values (close to zero) emerge, see Figures 2b,c, 3e and 4f. In particular as regards the interesting case of the feature `x_acf1` (i.e., the sample autocorrelation at the lag equal to 1), it seems that this takes either positive or negative values across the contiguous United States for precipitation at the 3-month temporal resolution (i.e., the seasonal time scale), with these values having a mean that is close to zero (see Figure 2b).

## 3.2 Spatial hydroclimatic patterns and feature importance

In Figures 7 and 8, we present the four groups of geographical locations obtained for each set {time series type, temporal resolution} through feature-based time series clustering. Mostly comparable spatial patterns across temporal resolutions emerge from our investigations, with geographical locations belonging to the same cluster at one temporal resolution also being grouped together at the remaining temporal resolutions, and rather fewer exceptions than confirmations of this rule. Concerning this finding, it is important to note that there were no prior expectations on whether it should/would emerge from the cluster analyses or not and, consequently, there are no explanations on why there are exceptions to it (e.g., the ones referring to the Mountain Time Zone; see Figures 7a,d,g,j,m and 8 a,d,g,j). Similarly, there are no initial expectations on whether the spatial patters should/would look alike between the present work and others that perform cluster analyses by using the same dataset (e.g., Kratzert et al. 2019; Jehn et al. 2020), as different characteristics have informed these latter analyses. Still, such similarities are also present, especially with the clusters proposed by Kratzert et al. (2019).

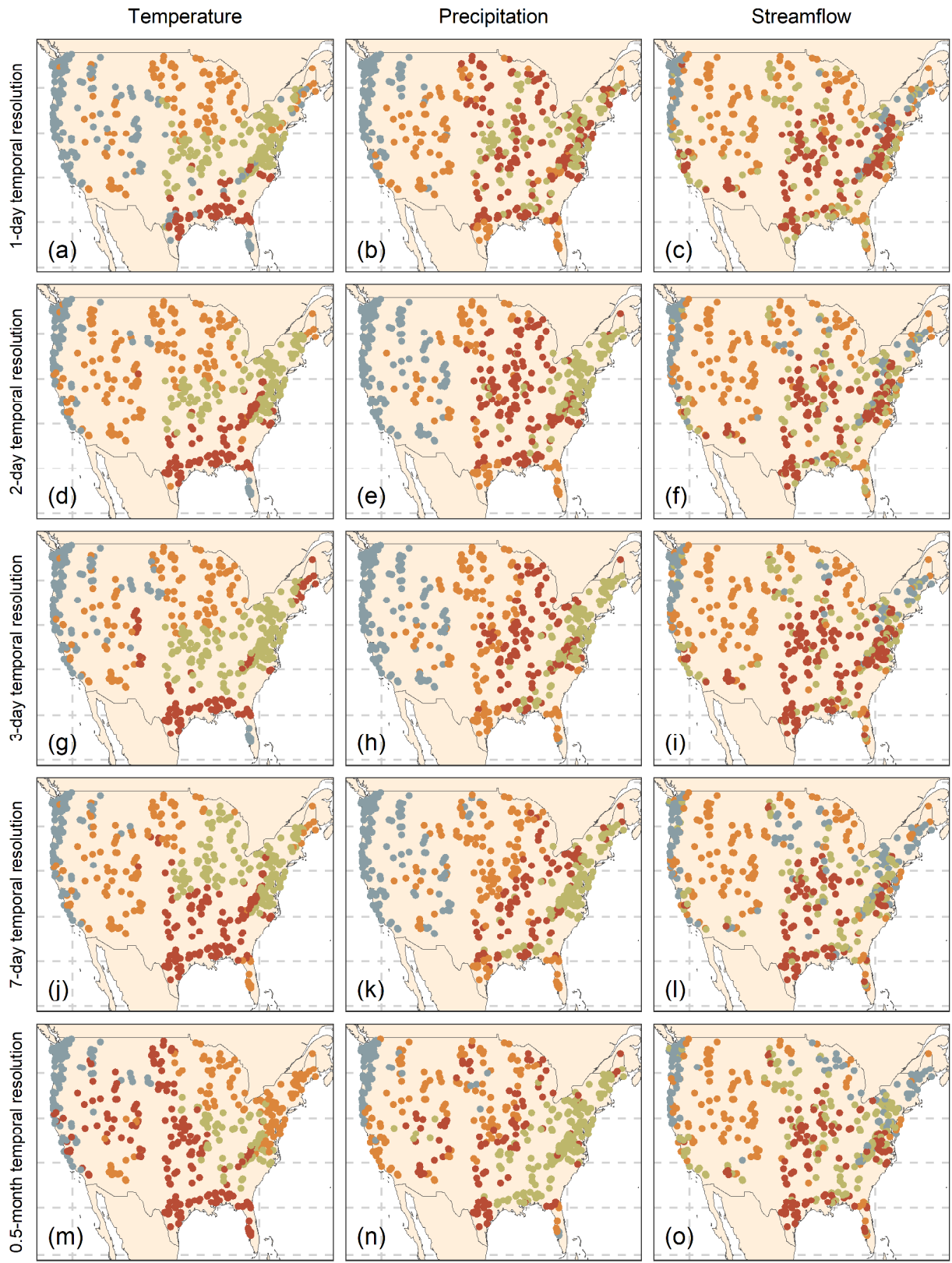

**Figure 7.** Clusters of (a, d, g, j, m) temperature, (b, e, h, k, n) precipitation and (c, f, i, l, o) streamflow at the (a–c) 1-day, (d–f) 2-day, (g–i) 3-day, (j–l) 7-day and (m–o) 0.5-month temporal resolutions. Clustering in four groups has been performed separately for each set {time series type, temporal resolution}. The same colours are used in all the panels; however, different ranges of feature values characterize the clusters of these panels.

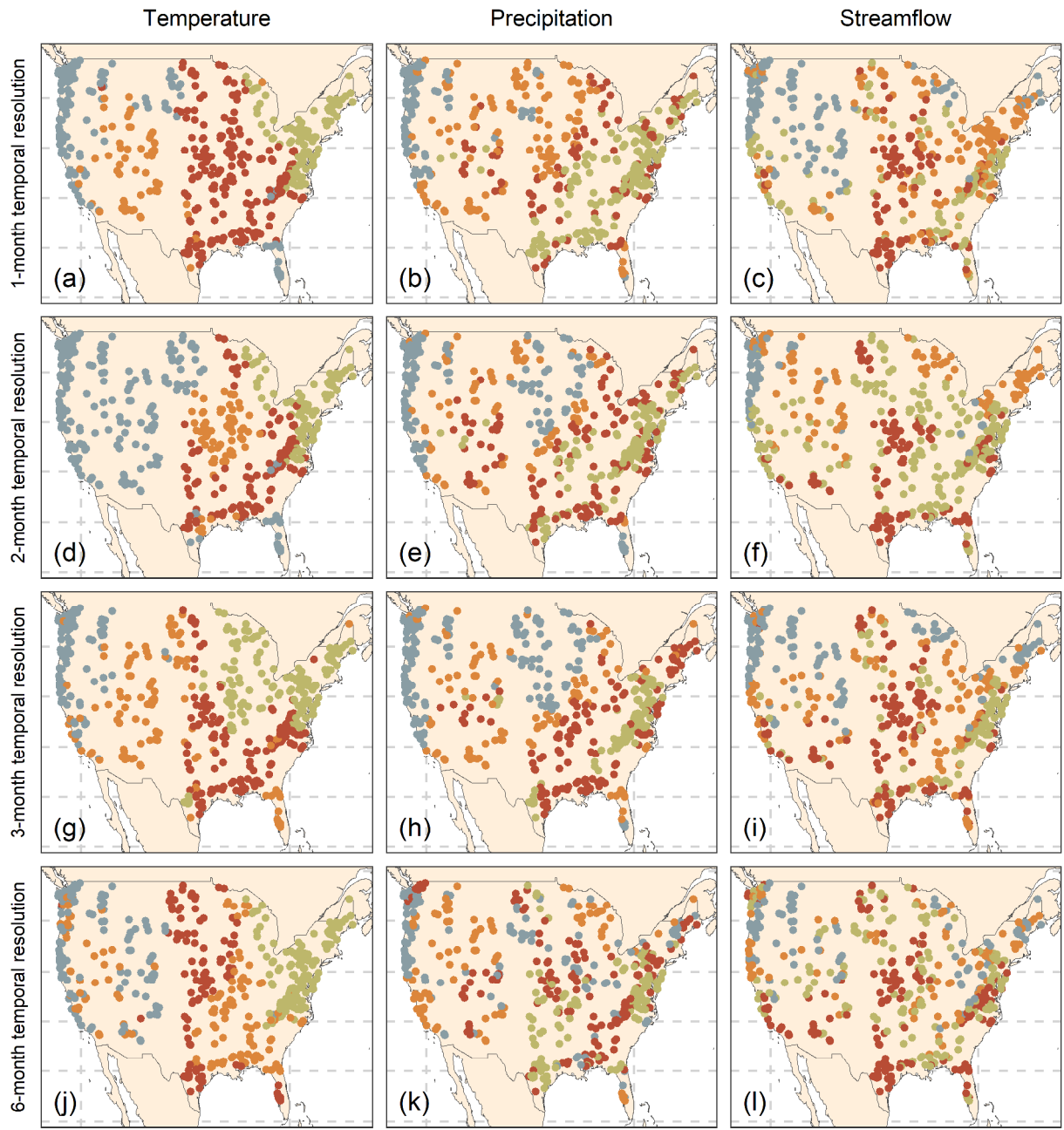

Figure 8. Clusters of (a, d, g, j) temperature, (b, e, h, k) precipitation and (c, f, i, l) streamflow at the (a–c) 1-month, (d–f) 2-month, (g–i) 3-month and (j–l) 6-month temporal resolutions. Clustering in four groups has been performed separately for each set {time series type, temporal resolution}. The same colours are used in all the panels; however, different ranges of feature values characterize the clusters of these panels.

More generally, we observe that neighbouring geographical locations tend to be attributed to the same cluster at all the examined temporal scales, especially for temperature and precipitation but for streamflow as well, although spatial information and information on the magnitude of the time series has not been considered in the clustering. This outcome rather suggests the relevance of the proposed feature compilation for multi-scale hydroclimatic time series analysis. Additionally, we observe that, especially for temperature and precipitation, the hydroclimatic clusters seem to be dividing, to a notable extent, the contiguous United States into "vertical sections". This fact could imply that the longitude is a more important factor than the latitude for determining hydroclimatic similarity in this continental-scale region.

A reasonable question that arises, at this point, is how reliable the above-summarized spatial hydroclimatic patterns are. To provide an answer to this question, we examine the average silhouette widths for the cluster analyses that supported the extraction of these patterns. These widths are shown in Figure 9 (see the columns referring to four clusters) and are all quite larger than 0. They are also of similar magnitude as those referring to the optimal numbers of clusters for the contiguous United States according to Kratzert et al. (2019). Therefore, our cluster analyses and the patterns stemming from them are reliable according to the average silhouette width. However, according to the same metric, for all the datasets clustered in the present work, the optimal number of clusters is two.

Of course, this does not mean, at any chance, that dividing the datasets into four clusters and discussing the patterns emerging from such divisions is not meaningful, especially if we consider the pronounced spatial coherence of the clustering outcomes despite the fact that no spatial information and no information on the time series magnitude is considered in the clustering. Moreover, the optimal numbers of clusters could be different in terms of other metrics, which are outside of the scope of the present work. Still, it is also highly relevant to inspect the clustering outcomes for the case in which the optimal number of clusters according to Figure 9 is adopted (see Figures 10 and 11). Notably, spatially coherent patterns emerge in this case as well. These patterns are again largely comparable across time series types and across temporal resolutions.

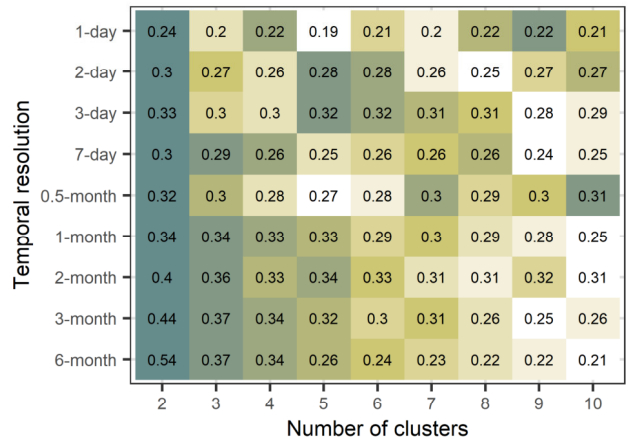

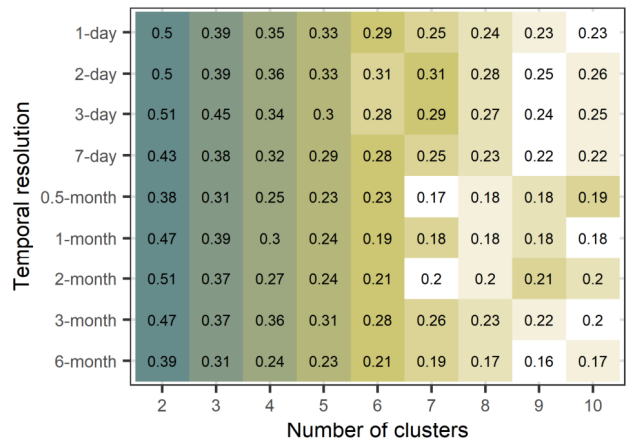

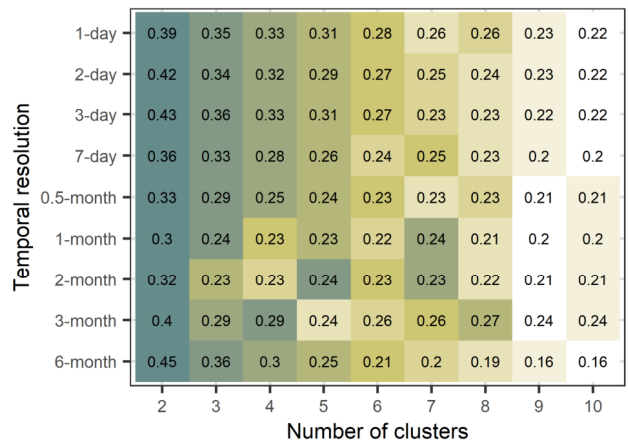

Figure 9. Average silhouette widths computed for assessing the clustering analyses of the (a) temperature, (b) precipitation and (c) streamflow time series at the nine examined temporal resolutions and for nine different numbers of clusters. The latter have also been ranked from the best (1st) to the worst (9th) in terms of average silhouette width.

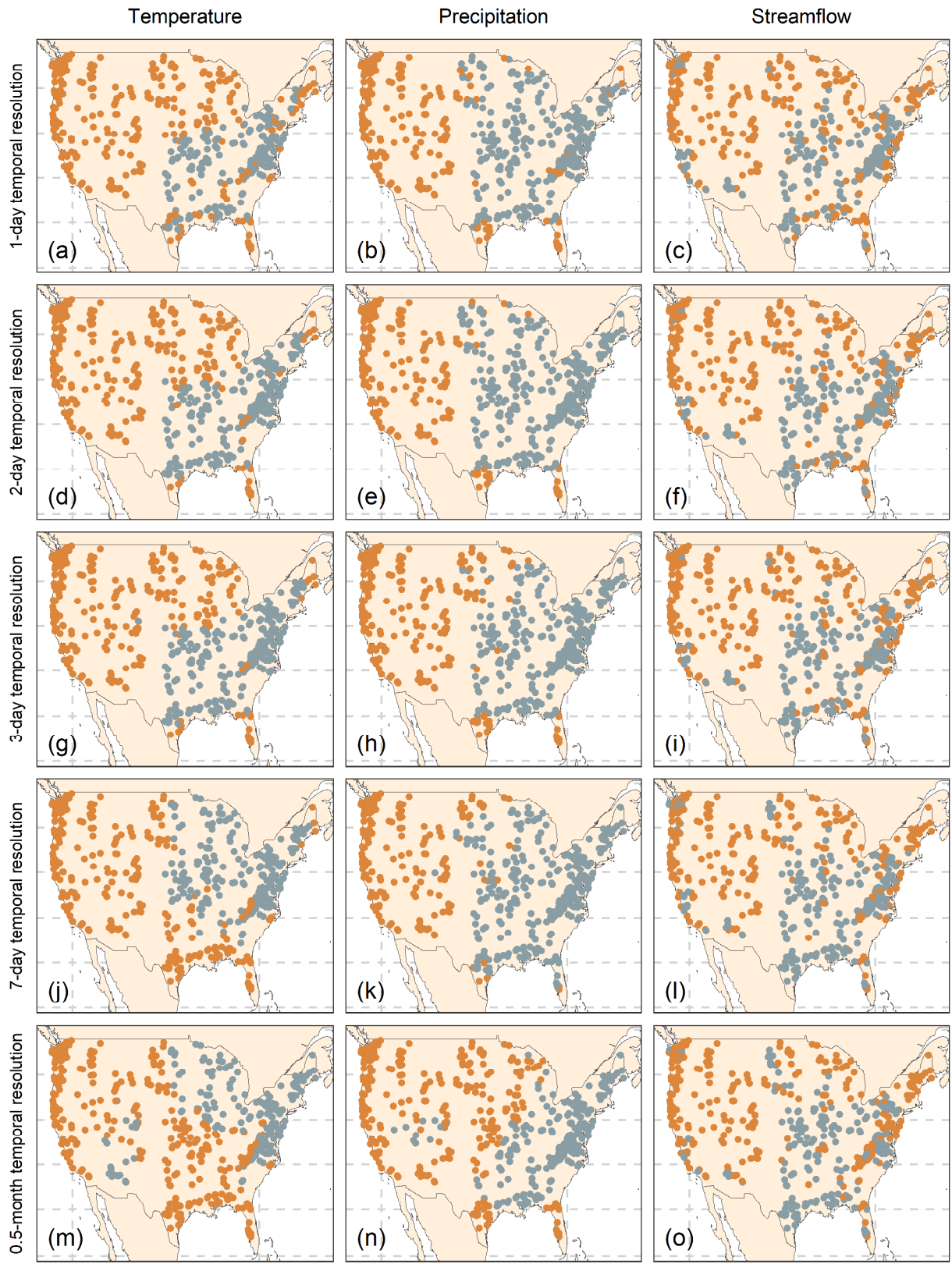

Figure 10. Clusters of (a, d, g, j, m) temperature, (b, e, h, k, n) precipitation and (c, f, i, l, o) streamflow at the (a–c) 1-day, (d–f) 2-day, (g–i) 3-day, (j–l) 7-day and (m–o) 0.5-month temporal resolutions. Clustering in two groups has been performed separately for each set {time series type, temporal resolution}. The same colours are used in all the panels; however, different ranges of feature values characterize the clusters of these panels.

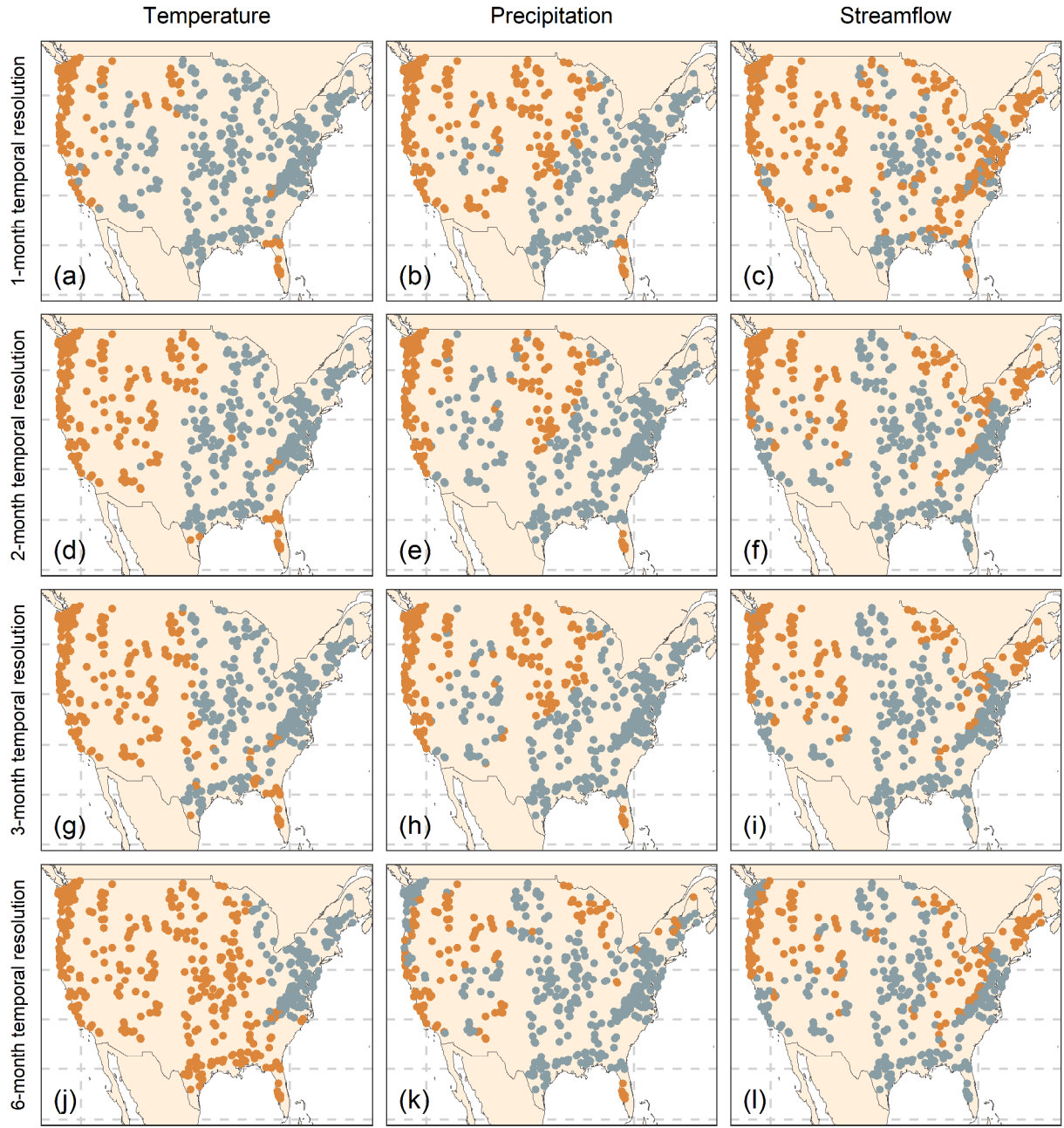

Figure 11. Clusters of (a, d, g, j) temperature, (b, e, h, k) precipitation and (c, f, I, l) streamflow at the (a–c) 1-month, (d–f) 2-month, (g–i) 3-month and (j–l) 6-month temporal resolutions. Clustering in two groups has been performed separately for each set {time series type, temporal resolution}. The same colours are used in all the panels; however, different ranges of feature values characterize the clusters of these panels.

Furthermore, the rankings of the features according to their usefulness-importance in time series clustering are presented in Figure 12. For most of the features (specifically, for the features `x_acf1`, `x_acf10`, `diff1_acf1`, `diff1_acf10`, `diff2_acf1`, `diff2_acf10`, `seas_acf1`, `x_pacf5`, `diff1x_pacf5`, `diff2x_pacf5`, `seas_pacf`, `std1st_der`, `entropy`, `stability`, `spike`, `e_acf1`, `e_acf10` and

`seasonal_strength`), these rankings can differ across temporal resolutions and/or time series types, thereby pointing out the usefulness of massive feature extraction for the study of hydroclimatic similarity (especially in cases when no prior information is available at hand about the most important features for the time scale of interest). A question that could naturally arise, at this point, is whether the differences across time scales and time series types are a consequence of high correlation among some of the features, making the clustering algorithm to pick the one or the other interchangeably. The answer to this question is negative, as random forests are known to be skillful in handling highly correlated features (Boulesteix et al. 2012; Ziegler and König 2014).

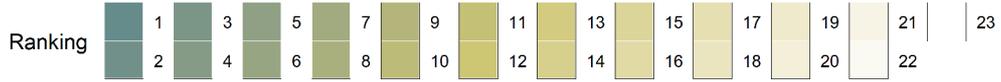

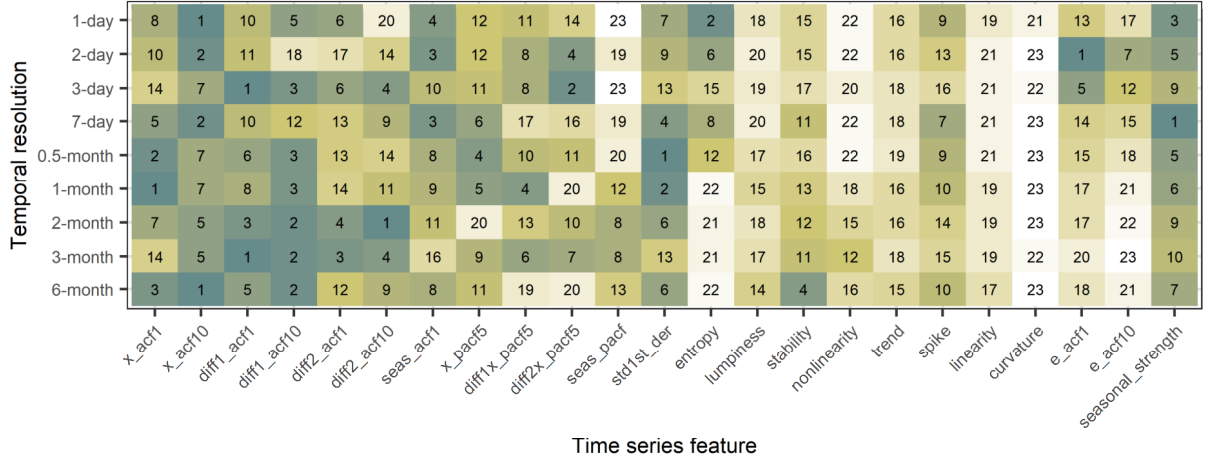

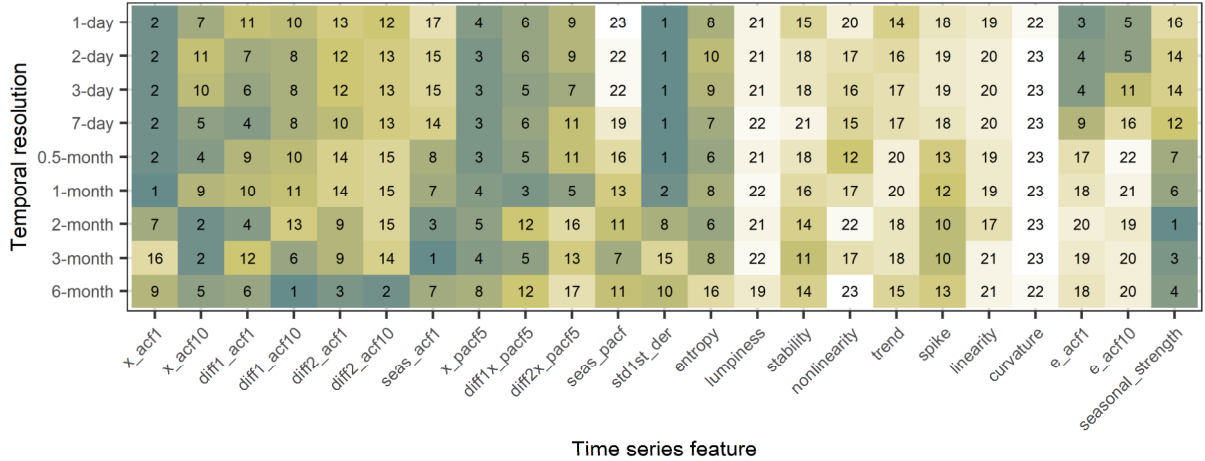

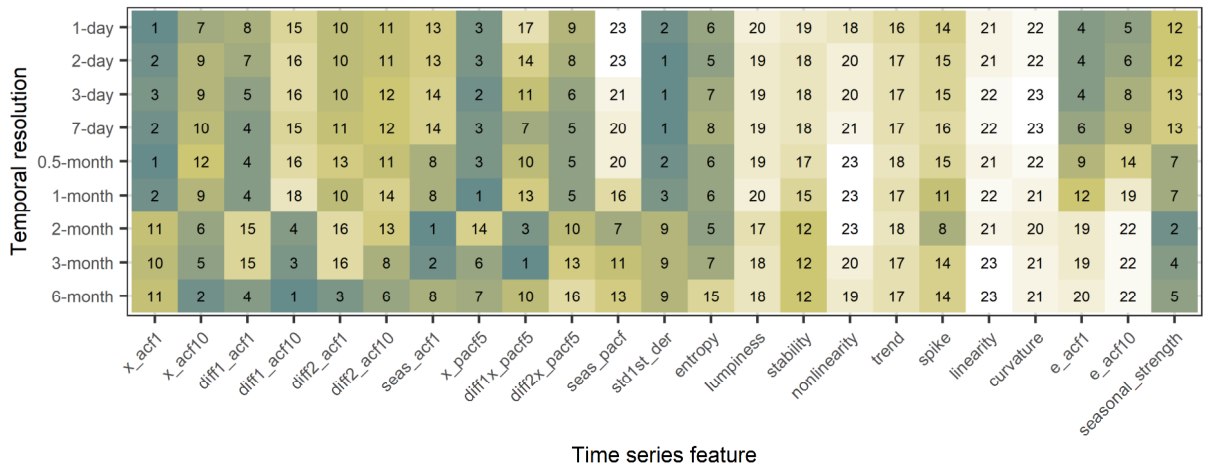

Figure 12. Rankings of the (a) temperature, (b) precipitation and (c) streamflow features from the most important (1st) to the least important (23rd) ones in feature-based time series clustering using random forests with 5 000 trees at the various examined temporal resolutions. The features are defined in Section 2.1.

Other abilities that make random forests appealing for clustering in general, and clustering upon numerous time series features in particular, are those of demonstrating high performance compared to other algorithms, operating successfully when interactions are present and being invariant to monotone transformations of the features (see, e.g., the review by Tyralis et al. 2019, Section 2.8.1). We believe that these properties justify the robustness of the results of the feature importance investigations, given also that the latter have been performed on a large dataset. Aside from the above-mentioned features, the remaining ones (i.e., `lumpiness`, `nonlinearly`, `linearity`, `trend` and `curvature`) are mostly among the least important ones independently of the temporal scale and time series type. Also notably, for precipitation and streamflow time series, there is a clear distinction in terms of feature importance between the time scales that are coarser than monthly and the remaining ones, possibly suggesting substantial differences in the characteristics between these two groups of time scales.

## 4. Further discussions

Overall, the proposed feature compilation has facilitated efficient investigations "at scale" of the behaviours of hydroclimatic processes, with the term "at scale" implying several notions of scale according to Taylor and Letham (2018). The most characteristic of these notions are related to the essential requirements of collectively studying: (a) large numbers of time series; (b) time series with varying temporal resolutions; (c) time series with varying lengths; (d) different types of time series; and (e) time series representing various hydroclimatic (and catchment) conditions. Feature investigations, conducted for temperature, precipitation and streamflow processes in terms of temporal dependence, temporal variation, "forecastability", lumpiness, stability, nonlinearity (and linearity), trends, spikiness, curvature and seasonality, have offered characterizations that have been missing from the previous literature. We believe that this is an important contribution, given the relevance of multi-scale time series analysis to deepening our perception of hydroclimatic behaviours. Indeed, we truly understand a hydroclimatic process only if we have studied its behaviour at multiple, and as many as possible, temporal scales (see the relevant discussions in McKitrick and Christy 2019). Moreover, the proposed feature compilation could effectively support the assessment of stochastic simulation methods in hydrology (see, e.g., the multi-scale assessment in Koutsoyiannis and Onof 2001), given also its appropriateness for utilization "at scale" and the relevance of its features for conducting hydroclimatic research. For a thorough justification of this relevance, the reader is referred to the overview by Papacharalampous et al. (2021).

Regarding the large-scale results of this work, these have shed new light into the similarities and differences between temperature, precipitation and streamflow processes. Altogether, we have found that the similarities are quite pronounced, as also suggested by previous large-scale investigations characterizing different aspects of the structure of hydroclimatic time series (Dimitriadis et al. 2021; Papacharalampous et al. 2021, 2022). Indeed, temperature, precipitation and streamflow processes have been herein found to exhibit quite analogous evolution patterns of their characteristics with increasing (or decreasing) temporal resolution in terms of many of the examined time series features (see again Figures 2–6). This form of similarity is somewhat more intense between precipitation and streamflow processes, and sometimes also between temperature and precipitation processes (probably because of the physical relationships between each of these two pairs, especially between precipitation and streamflow). Of course, this does not mean, by any chance, that we have not identified notable differences between the examined time series types as regards the evolution patterns of their features with increasing (or decreasing) temporal resolution nor that their features are similar at the examined temporal resolutions (as, in fact, they are not).

Notable similarities and differences have also been identified as regards the spatial variability of the temperature, precipitation and streamflow features at the examined temporal resolutions, as well as the usefulness-importance of the features in feature-based time series clustering using random forests. It seems that, although the spatial patterns obtained for each of the examined time series types are rather analogous across temporal resolutions, the most important features in grouping together hydroclimatic time series (and, by extension, geographical locations) with similar characteristics can vary from temporal scale to temporal scale and from hydroclimatic time series type to hydroclimatic time series type. This might be suggesting that hydroclimatic similarity (or dissimilarity) should always be assessed using a large variety of features. Hydroclimatic similarity investigations can support the regionalization of models for technical and operational purposes (see, e.g., Guo et al. 2021; Wang et al. 2021).

We believe that the selected large dataset represents well the various hydroclimatic (and catchment) behaviours, given the closeness of the feature values computed for the 1-month temporal resolution to those previously computed using global datasets for the same temporal resolution by Papacharalampous et al. (2021). Still, the present work is not limitation-free and, therefore, its limitations and possible extensions-expansions of its methodological framework should be attentively discussed. Indeed, the present work has investigated hydroclimatic phenomena in a data-driven way aspiring to contribute to the general efforts made in the literature towards an improved understanding of these phenomena. However, additional in depth and in breadth investigations are needed to address puzzling scientific questions such as *"Why temperature is more forecastable-predictable at the time scale x than at the time scale y?"*. For instance, the study by Klemeš (1974) entitled *"The Hurst phenomenon: A puzzle?"* explains that in depth explanations of real-world phenomena are hard to provide by looking at estimated statistics.

Also notably, since the catchments of the selected large dataset are small or medium and devoid of human intervention, the hydrographs are expected to respond quickly to precipitation. Indeed, the similarity between precipitation and streamflow processes as regards the evolution patterns of their features with increasing (or decreasing) temporal resolution could appear somewhat less intense, if time series originating from both small-medium and large catchments had been analysed instead. Notably, the proposed methodological framework can directly facilitate such analyses. Nonetheless, in its present form, it does not consider possible abrupt changes in the feature values possibly caused by human intervention (which do not occur in the examined dataset), as the features are computed once for each time series using its entire length. For catchments that have been modified during the data collection period, it is suggested that the features are computed separately before and after the change in time periods were stationarity is a befitting assumption.

Perhaps even more importantly, the temporal resolution and lengths of the original time series selected as a starting point for our analyses have allowed robust investigations for temporal resolutions from the 1-day and up to the 6-month ones, while the study of larger and smaller temporal resolutions is also essential for obtaining a deep understanding of hydroclimatic behaviours. In this view, a direct extension of this work would require the computation of the features for time series with finer temporal scales (e.g., the hourly or even sub-hourly ones) and their aggregations up to the daily temporal scale. In a similar vein, longer daily records (see e.g., the time series lengths in McKitrick and Christy 2019) or even paleoclimatic time series could be exploited to extend the analyses of this work to temporal scales larger than the 6-month one (see, e.g., the large span of temporal scales investigated in Markonis and Koutsoyiannis 2013). Moreover, spatial correlations could have an impact on our results, given the spatial heterogeneity characterizing the selected dataset. In fact, some regions have higher density of sampling points than others. This impact could be the subject of future research.

Possible expansions of the proposed framework could further include spatial pattern identification for other continental-scale regions around the globe or even at the global scale, followed by investigations on the existence of possible similarities or dissimilarities in the identified spatial patterns across various temporal scales. Lastly, the possibility of predicting hydroclimatic time series features at fine temporal scales by using the values of hydroclimatic time series features at larger temporal scales within regression-based regionalization frameworks (see, e.g., the ones designed and studied by Tyralis et al. 2019b, 2021b and Papacharalampous and Tyralis 2022 for providing point and quantile predictions of features and signatures) could be investigated for improving the quality of feature-based time series simulations at geographical locations lacking observations at fine temporal resolutions.

Statistical and machine learning algorithms (see, e.g., the comprehensive lists and descriptions provided by Hastie et al. 2009; James et al. 2013), and especially random forecasts that are not affected by redundant predictors by construction (Tyralis et al. 2019a) and the various boosting algorithms that perform intrinsic variable selection (Tyralis and Papacharalampous 2021), could be merged with the feature compilation proposed in this work for fulfilling (some of) the aforementioned tasks (as it has been made herein for spatial pattern identification and feature importance computation; see also Yang and Chui 2021). The importance of exploiting diverse concepts and methods from various scientific fields in earth system science has been emphasized by Moallemi et al. (2021).

## 5. Conclusions

We have proposed and thoroughly applied a new methodological framework for the study of the cross-scale properties of hydroclimatic time series. Starting from mostly larger feature compilations sourced from the data science literature (Wang et al. 2006; Fulcher et al. 2013; Hyndman et al. 2015; Kang et al. 2017, 2020) and, at the same time, exploiting the results of previous large-sample investigations on a single temporal resolution from the hydroclimatic literature (Papacharalampous et al. 2021, 2022), we have composed a feature compilation for multifaceted and automatic hydroclimatic time series analysis at multiple temporal scales. This compilation is the core of the proposed framework and includes a variety of interpretable features, whose computation requires minimal (practically zero) adaptations across temporal scales, and whose values do not depend on the length of the time series.

Based on this specific compilation, extensive feature investigations and comparisons across nine temporal resolutions (i.e., the 1-day, 2-day, 3-day, 7-day, 0.5-month, 1-month, 2-month, 3-month and 6-month ones) and three time series types (i.e., temperature, precipitation and streamflow) have been conducted over the contiguous United States. Detailed characterizations have been obtained (which are particularly important –from a practical perspective– for the 1-day, 1-month and 3-month temporal scale, which are most commonly referred to as the "daily", "monthly" and "seasonal" ones, respectively), and various similarities and differences between the examined time series types have been identified. Furthermore, feature-based clustering has been performed for investigating the spatial variability of the temperature, precipitation and streamflow features across the examined continental-scale region and across temporal resolutions, and the significance of using a variety of features for the assessment of hydroclimatic similarity has been discussed based on results obtained through explainable machine learning. To the authors' knowledge, both the clustering analyses at multiple temporal resolutions based on a large variety of features and the cross-scale explainable machine learning explorations on the relative importance of the features in the clustering are new approaches to studying hydroclimatic similarity.

We believe that our contribution to the literature is important, especially as regards the previously missing characterizations and the similarities identified across time series types in the evolution patterns of their features with increasing (or decreasing) temporal resolution, given the substantial relevance of multi-scale time series analysis to improving our understanding of hydroclimate (McKitrick and Christy 2019) and the utility of the same analysis in assessing time series simulation methods, especially the time series disaggregation ones (see, e.g., the assessment in Koutsoyiannis and Onof 2001). Future research could, among others, extent the investigations of this work to other temporal resolutions and time series types. It could also compute the features of this work to assess a variety of time series disaggregation methods.

**Acknowledgements:** We are sincerely grateful to the Associate Editor and the Reviewers for their constructive suggestions. The first preprint version of this work was featured in the newsletter «This Week in A.I.» by DeepAI (https://deepai.org/this-week-in-ai) on 4 December 2021, together with four other works, as the "week's most popular data science and artificial intelligence research". YM and MH were supported by the project "Investigation of Terrestrial HydrologicAl Cycle Acceleration (ITHACA)" funded by the Czech Science Foundation (Grant 22-33266M).

**Appendix A    Information on statistical software and data availability**

The analyses and visualizations have been performed in R Programming Language (R Core Team 2021). The following contributed R packages have been used: cluster (Maechler et al. 2021), cowplot (Wilke 2020), data.table (Dowle and Srinivasan 2021), devtools (Wickham et al. 2021), gridExtra (Auguie 2017), knitr (Xie 2014, 2015, 2021), randomForest (Liaw and Wiener 2002; Liaw 2018), rmarkdown (Xie et al. 2018; Xie et al. 2020; Allaire et al. 2021), stringi (Gagolewski 2021), tidyverse (Wickham et al. 2019; Wickham 2021), tsfeatures (Hyndman et al. 2020), wesanderson (Ram and Wickham 2018) and zoo (Zeileis and Grothendieck 2005; Zeileis et al. 2021). The original data can be retrieved from Newman et al. (2014) and Addor et al. (2017a). The R codes for multi-scale time series feature estimation and cluster analyses will be made available upon request to the corresponding author.

**Appendix B    Supplementary material**

Supplementary material can be found in the online version of the paper and provides basic definitions for supporting the understanding of the adopted time series feature estimation process, along with an extended version of Section 3. This version includes additional visualizations, which present: (a) the rankings of the feature values obtained for the 1-day, 2-day, 3-day, 7-day, 0.5-month, 1-month, 2-month, 3-month and 6-month temporal resolutions, separately for each set {time series feature, time series type, geographical location} (Figures S6–S28); (b) the side-by-side boxplots of the feature values characterizing the four groups of geographical locations obtained for each set {time series type, temporal resolution} through feature-based time series clustering (Figures S31–S57); (c) the side-by-side boxplots of the feature values characterizing the two groups of geographical locations obtained for each set {time series type, temporal

resolution} through feature-based time series clustering (Figures S61–S87); (d) the rankings of the features from the most to the least important ones, separately for each set {time series type, temporal resolution}, in feature-based time series clustering using random forests with 500, 1 000, 2 000, 3 000 and 4 000 trees (Figures S89–S93); and (e) the Pearson correlations between hydroclimatic time series features at the 1-day, 2-day, 3-day, 7-day, 0.5-month, 1-month, 2-month, 3-month and 6-month temporal resolutions (Figures S94–S147).

**Author contributions:** GP and HT conceptualized the work and designed the experiments with input from YM and MH. GP performed the analyses and visualizations, and wrote the first draft, which was commented and enriched with new text, interpretations and discussions by HT, YM and MH.